\documentstyle{article}
\title{\bf Radiation Field on Superspace$^{\dagger}$}
\author{ Pedro F. Gonz\'alez-D\'{\i}az.\\
Instituto de Matem\'aticas y F\'{\i}sica Fundamental\\
Consejo Superior de Investigaciones Cientificas\\
Serrano 121, 28006 Madrid (SPAIN)\\
}
\date{June 8, 1994}
\begin{document}
\maketitle
\large
\setlength{\baselineskip}{0.5cm}
\vspace{3cm}


We study the dynamics of multiwormhole configurations within the
framework of the Euclidean Polyakov approach to string theory,
incorporating a modification to the Hamiltonian which
leads to a Planckian probability measure for the
Coleman parameters $\alpha$ that allows
$\frac{1}{2}\alpha^{2}$ to be interpreted as the energy of the quanta of
a radiation field on superspace whose values might still
fix the coupling constants.


\vspace{6.5cm}

\noindent $^{\dagger}$To appear in {\it Geometry of Constrained Dynamical
Systems}, ed. J.M. Charap (Cambridge Univ. Press, Cambridge, 1995).


\pagebreak

Multiwormhole configurations in Polyakov string theory have been studied
by looking at the wormholes as the handles on a Riemann surface of
genus $g$, with $g$ giving the number of handles or wormholes in the
configuration [1]. The Green function that describes the effects of such
wormholes on first order tachyonic amplitudes was calculated by
Lyons and Hawking [2] as a path integral over all space-time coordinates
$x_{\mu}$ on the Riemann surface. It was assumed that
all the fields have the
same values at the points on the two circles which result after
cutting the handles in such a way that they become divided
in two topologically separated discs. In this case, the points were
identified by the projective transformations of the Schottky group
on each pair of circles, and the Green function can be written as [1-3]
\begin{equation}
<x_{\mu}(z_{1}).x_{\nu}(z_{2})...>=\int
d[x_{\mu}]x_{\mu}(z_{1})x_{\nu}(z_{2})...\prod_{r}\prod_{n}\delta(x_{n}-x'_{n})e^{-I},
\end{equation}
where $I$ is the Euclidean action, $r$ runs from 1 to $g$,
\begin{equation}
x=\sum x_{n}e^{i\zeta n},
\end{equation}
and the delta function ensures that $\zeta$ on one circle is identified
with $\zeta '$ on another. On can express the Green function (1) in
terms of the handle quantum state on the circles
using the Fourier transform of the delta
function for the zero mode, and expanding the delta function for the
nonzero modes in terms of the complete set of orthonormal harmonic-oscillator
eigenstates which are the solutions of the string analogue
of the Wheeler DeWitt equation [2]
\begin{equation}
H_{WDW}\Psi_{nm_{n}^{(i)}}=[-\frac{\partial^{2}}{\partial
x_{0}^{2}}+\frac{1}{2}\sum_{n>0,i}(-\frac{\partial^{2}}{\partial(Y_{n}^{(i)})^{2}}+n^{2}(Y_{n}^{(i)})^{2})]\Psi_{nm_{n}^{(i)}}=0,
\end{equation}
with solution ($i=1,2$)
\begin{equation}
\Psi_{nm_{n}^{(i)}}\propto
e^{-\frac{1}{2}n(Y_{n}^{(i)})^{2}}H_{m_{n}^{(i)}}(n^{\frac{1}{2}}Y_{n}^{(i)})\Psi_{K}(x_{0}), \Psi_{K}(x_{0})=e^{iK.x_{0}}.
\end{equation}
Eq. (3) is the canonically quantised version of the Hamiltonian constraint
derived [2] from the string Euclidean action for the field $x$
on the region of the complex plane outside a disc of radius
$r=-\frac{\ln\mid z\mid}{t}$ ($t$ is some Euclidean time), $K$
is the momentum of the zero mode $n=0$, and
\[Y_{n}^{(1)}=\frac{1}{2}(x_{n}+x_{-n}),
Y_{n}^{(2)}=\frac{1}{2i}(x_{n}-x_{-n}).\]

Now, as in the 4-dimensional case [4], one can calculate the
effect on Green functions in the fundamental region, by
doing a path integral over all fields $x^{\mu}$ on the
complex plane with the boundary conditions on the circles
given by a set of values $Y_{n}^{(i)}$, weighting
with the wave function $\Psi_{nm_{n}}^{(i)}$. Again as in
the space-time wormhole case, the effect will be given
by a vertex operator located approximately
at the center of the circles.
One can see that, after integrating over the fields, the
resulting path integral contains a factor $\kappa=(K^{2}+\sum\mid n\mid
m_{n}^{(i)}-2)^{-1}$.
We obtain thus a bi-local effective action [1]
\[-\int d\sigma_{1}\int d\sigma_{2}\sum\int d^{4}K\kappa
V_{p}(\sigma_{1})V_{p}(\sigma_{2}),\]
where the $V_{p}$ are the handle vertex operators for
the fields on the two circles. One can again convert the
bi-local action to a local action, by introducing $\alpha$
parameters, to finally obtain a probability measure with the
same general form as for the 4-dimensional case [5]. Nevertheless,
since now the $\alpha$ parameters are labelled by momentum $K$,
and also the occupation numbers $m_{n}^{(i)}$ for modes $n>0$,
these parameters
should be interpreted [1,3] as a quantum field on the infinite
dimensional superspace of all coordinate field values on a single circle.
Such a quantum field would be regarded as an infinite tower of fields on
the usual space-time minisuperspace; each stage along this tower is
labelled by the quantum number $m_{n}^{(i)}$ that defines the excited states of
the
basis set of solutions (4). In light of this interpretation, it was
concluded [1,3] that the $\alpha$'s could not be regarded as a set of coupling
constants to be fixed by quantum measurement, as required by the
Coleman mechanism to fix the coupling constants [6].

However, it is not quite clear that this conclusion can be maintained
because
in string theory the
Hamiltonian should be modified [7] by the addition of an (infinite) constant,
in such a way that just the ground state of the harmonic oscillators,
multiplied by the wave function of the zero mode, will obey the
Hamiltonian constraint.

We shall follow here a modified approach where this
shortcoming is avoided. The idea
consists in considering the more general case allowing for
handles on the Riemann surface
which, under cutting, give rise to
pairs of discs that are no longer disconnected to each other [8]. This
would ultimately imply partial or total breakdown of the Schottky
group invariance under projective transformations between discs [9].
Such handles need not be
on shell in the sense that the analogue of the Wheeler DeWitt operator
acting on the excited eigenstates $\Psi_{m_{n}^{(i)}}(Y_{n}^{(i)})$ is no
longer zero, but gives the corresponding harmonic-oscillator eigenvalues
$nm_{n}^{(i)}$ [8,10]. In this case, the $\delta$ function for all field modes
in the Green function should be replaced [8] by a path integral which
has the $x_{n}$ fixed at the given values at time $t=0$ on a circle
and $t=t_{1}$ on another, for some Euclidean time interval $t_{1}$
between the two circles. Unlike for the states given by
Eqns. (1)-(3), this path
integral is not generally separable into a product of wave
functions (4), but, after integrating over time $t_{1}$, it gives
the density matrix for a mixed quantum state,
and is equal to the
propagator
\begin{equation}
K(x_{n},0;x'_{n},t_{1})=\prod_{n>0,i}\sum_{m_{n}^{(i)}>0}\Phi_{m_{n}^{(i)}}(Y_{n}^{(i)})\Phi_{m_{n}^{(i)}}(Y_{n}^{(i)'})e^{-nm_{n}^{(i)}t_{1}},
\end{equation}
where $\Phi_{m_{n}^{(i)}}(Y_{n}^{(i)})$ would match the excited
eigenstates of the harmonic oscillators ($\Phi_{m_{n}^{(i)}}$
$\equiv\Psi_{m_{n}^{(i)}}$) if we allowed
an unlimited resolution for the fields, or equal some wave functions which
contain only that part of the information contained in the $\Psi_{m_{n}^{(i)}}$
that associates with the finite eigenenergy sector surviving
after the introduction of a given cut off at a finite highest
energy scale.
If the handles are off shell, then,
instead of the quantum constraint equation (3),
we must use the "time-independent" wave
equation [8,10]
\begin{equation}
(H_{WDW}-\sum_{n>0,i}nm_{n}^{(i)})\Psi=0,
\end{equation}
so that the wave function for the wormhole handles becomes
\begin{equation}
\Psi\equiv\Psi(x_{0},Y)=e^{iKx_{0}}\prod_{n>0,i}e^{-\frac{1}{2}n(Y_{n}^{(i)})^{2}}.
\end{equation}

The $K^{2}$ term in (6), resulting from the application of operator
$-\frac{\partial^{2}}{\partial x_{0}^{2}}$ to $\Psi$,
gives the energy of the $x_{0}$ plane waves, with
$K^{2}<0$. This would correspond to a timelike momentum in {\it Lorentzian}
space-time. Wick rotating to Euclidean momentum, $K\rightarrow -iK_{E}$,
we have from (6) and (7)
$K_{E}^{2}=\sum_{n>0,i}(m_{n}^{(i)}+\frac{1}{2})n$
and, since we are
dealing with harmonic oscillators, we have in general
$K_{E}=\sum_{n>0,i}m_{n}^{(i)}n$ [11].
Let us then calculate the quantity
\[\sum_{m_{n}^{(i)}>0}\Psi(x_{0},Y)\Psi^{*}(x_{0}',Y')\]
\begin{equation}
=\sum_{m_{n}^{(i)}>0}\prod_{n>0,i}(e^{-\frac{1}{2}n(Y_{n}^{(i)})^{2}}e^{-\frac{1}{2}n(Y_{n}^{(i)'})^{2}}e^{-m_{n}^{(i)}n(x_{0}'-x_{0})}).
\end{equation}
The relative minus sign between $x_{0}$ and $x_{0}'$ in (8) should be kept
anyway in order to ensure an orientable surface when the handles are glued
together [2]. Taking $x_{0}'-x_{0}=t_{1}$, noting that, since each
two circles can have any time separations,
one should integrate (8) over
all possible values of $t_{1}$ [8], and denoting the density matrix
by $\rho$,
we can see that
\[i\int dt_{1}K(x_{n},0;x_{n}',t_{1})\]
\begin{equation}
=\sum_{m_{n}^{(i)}>0}\int d(x_{0}'-x_{0})\Psi(x_{0},Y)\Psi^{*}(x_{0}',Y')\equiv
i\rho(Y;Y'),
\end{equation}
whenever we take for the states $\Phi_{m_{n}^{(i)}}$ in the propagator (5) only
that
part of the $\Psi_{m_{n}^{(i)}}$ which corresponds to the harmonic oscillator
ground states, surviving after projecting off all the information
contained in the Hermite polynomials.

{}From (7) and (9), the density matrix for handles becomes
\begin{equation}
\rho(Y;Y')=\sum_{m_{n}^{(i)}>0}\prod_{n>0,i}\frac{\Psi_{0}(Y_{n}^{(i)})\Psi_{0}(Y_{n}^{(i)'})}{nm_{n}^{(i)}},
\end{equation}
where $\Psi_{0}(Y_{n}^{(i)})=e^{-\frac{1}{2}n(Y_{n}^{(i)})^{2}}$.
Thus, for each $m_{n}^{(i)}>0$,
we should use a Green function for
the density matrix of the mixed state case given by

\[<x_{\mu}(z_{1}).x_{\nu}(z_{2})...>\]
\begin{equation}
=\int
d[x_{\mu}]x_{\mu}(z_{1})x_{\nu}(z_{2})...\prod_{r}\prod_{n>0,i}\tilde{\Theta}(x_{0}-x_{0}')\Psi_{0}(Y_{n}^{(i)})\Psi_{0}(Y_{n}^{(i)'})e^{-I},
\end{equation}
where we have replaced the full $\delta$ function in (1) for
the density matrix (7), specialising to a single generic
relative probability $(nm_{n}^{(i)})^{-1}$ for each mode
$n>0$, and the step function
$i\tilde{\Theta}(x_{0}-x_{0}')=\Theta(x_{0}-x_{0}')$ arises from
integrating the $\delta$ function for the zero mode over its
argument, as it is done in (9) and (10). In the present
approach, if we want to consider a Green function also for
the wave function (7), instead of the full $\delta$ function,
we should use the probability $\mid\Psi\mid^{2}$ obtained
from (7) as the weighting factor. This Green function will
then be
\[<x_{\mu}(z_{1}).x_{\nu}(z_{2})...>\]
\begin{equation}
=\int
d[x_{\mu}]x_{\mu}(z_{1})x_{\nu}(z_{2})...\prod_{r}\prod_{n>0,i}\delta(x_{0}-x_{0}')\Psi_{0}(Y_{n}^{(i)})\Psi_{0}(Y_{n}^{(i)'})e^{-I}.
\end{equation}

If we regard each pair of circles as the ends of a sum of wormholes of
different species [1,3], then each species would now be labelled by just the
momentum
$K$ of the zero mode, but {\it not} the levels $m_{n}^{(i)}$ of the other
modes. The
quantity $K$ can be interpreted as the conserved scalar charge carried
by the wormhole [7].

We can now calculate the effect of wormholes on tachyonic amplitudes for
handles
whose quantum state is given by both
a density matrix and a wave function,
using the
procedure devised by Lyons and Hawking [2].
For the case of handles in mixed state
and tachyons with momenta $p_{j}$,
unlike the pure-state case considered in Ref. 2,
the path integral describing the interaction cannot be
factorised into path integrals on each of the two circles [8].
Instead of the wave function (4), one should then introduce
as weighting factor the density matrix element
$\rho_{m_{n}}^{(i)}$
which corresponds to each relative probability
$\frac{1}{nm_{n}^{(i)}}$. In the limit of small circle radius
$r\rightarrow 0$, we then have for each of these density
matrix elements
\[D(\rho;p_{j})\propto\int
dx_{0}dx_{0}'(\prod_{n>0}dY_{n}^{(i)}dY_{n}^{(i)'})\rho_{m_{n}^{(i)}}\]
\[\times\int drr^{-3+(\sum_{1}^{M}p_{j})^{2}}\int[{\it
D}\zeta_{j}]\exp[i(x_{0}-x_{0}')\sum_{1}^{M}p_{j}]\]
\begin{equation}
\times\prod_{n>0}e^{\sum_{i}[-\frac{1}{2}n((Y_{n}^{(i)})^{2}+(Y_{n}^{(i)'})^{2})+ir^{n}k_{n}^{(i)}(Y_{n}^{(i)}-Y_{n}^{(i)'})]},
\end{equation}
where
\[\rho_{m_{n}^{(i)}}=\prod_{n>0}\frac{\Psi_{0}(Y_{n}^{(i)})\Psi_{0}(Y_{n}^{(i)'})}{nm_{n}^{(i)}},\]
and we take, as in [2],
$r=\mid z_{2}\mid^{-1}$, $\zeta_{j}=\frac{z_{j}}{z_{2}}$, $j=3,...,M$,
with $M$ the number of on-shell tachyon vertex operators inserted in the
region exterior to the circles, $\zeta_{0}=0$, $\zeta_{1}=\infty$,
$\zeta_{2}=1$. $\int[{\it D}\zeta_{j}]$ denotes integration over $\zeta_{j}$
with a measure whose explicit form need not be known for our calculation,
and $k_{n}=2\sum_{j=1}^{M}p_{j}\zeta_{j}^{-n}$, with $k_{n}^{(1)}$
and $k_{n}^{(2)}$ the real and imaginary parts of $k_{n}$, respectively.

Each integral pair over $Y_{n}^{(i)}$ and $Y_{n}^{(i)}$ gives a
factor $\frac{e^{-\frac{r^{2n}(k_{n}^{(i)})^{2}}{n}}}{n}$ for each $n$.
The Gaussian exponential factor would only contribute higher-order
interactions and will be disregarded in our calculation [2]. The other
possible contribution would come from integration over each pair of zero-mode
fields $x_{0}$,$x_{0}'$. Since all possible dependence of the density
matrix on such field has already been integrated out, unlike for handles
in a pure state,
we are left with
a single delta on $\sum_{j=1}^{M}p_{j}$, so in first order approximation
the path integral (13) gives essentially a factor
$\int\frac{dr}{r^{3}m_{n}^{(i)}n^{2}}$
for each $m_{n}^{(i)}$ and $n$.
Note that this factor contains no integration over momentum
$K$. As in the space-time wormhole case [4], the effect of
handles will be given again by a vertex operator, located
approximately at the center of the circles [1,3].
Thus, for each $m_{n}^{(i)}$ and $n$,
the stringy wormholes in mixed state will give rise
to a bi-local effective action [1] for each $m_{n}^{(i)}$
\begin{equation}
-\frac{1}{2}\int d\sigma_{1}\int
d\sigma_{2}\sum_{q,i}\frac{V_{q}(\sigma_{1})V_{q}(\sigma_{2})}{m_{n}^{(i)}n^{2}},
\end{equation}
where $V_{q}$ are the vertex functions. Following hereafter the same
procedure as for wormholes in space-time [4],
this action can be made local
by introducing $\alpha$ parameters, i.e.
\begin{equation}
\int
d\sigma\sum_{q,i}((-\frac{1}{2}\alpha_{q}^{2}m_{n}^{(i)}n^{2})+\alpha_{q}V_{q}(\sigma))
\end{equation}
which leads to a probability measure over the $\alpha$ parameters
for each $m_{n}^{(i)}$
\begin{equation}
Z(\alpha)\prod_{q,i}e^{-\frac{1}{2}\alpha_{q}^{2}m_{n}^{(i)}n^{2}},
\end{equation}
where again
$Z(\alpha)$ is the path integral over all fields $x^{\mu}$ on the
two-sphere [1,3],
containing the effective interaction $\alpha_{q}V_{q}(\sigma)$.
The distribution for $\alpha$-parameters associated with (16)
corresponds to just one of the infinite relative probabilities
for the state $\Psi_{0}(Y_{n}^{(i)})$ of handles. Therefore,
one should now sum (16) over all
$m_{n}^{(i)}$ [12,13], to finally obtain a probability measure
\begin{equation}
Z(\alpha)\prod_{q}(e^{\frac{1}{2}n^{2}\alpha_{q}^{2}}-1)^{-1}
\end{equation}
for each $n$ and $i$. Thus, as it was suggested for
4-dimensional space-time [12,13],
we obtain a Planckian distribution for $\alpha$
parameters that allows to interpret $\frac{1}{2}\alpha_{q}^{2}$ as the
energy of the quanta of a radiation field, and $n^{-2}$ as some temperature,
on string-theory superspace.

In the case that the quantum state of the handles be given by the
wave function (7),
the calculation is similar, but with
$\rho_{m_{n}^{(i)}}$
replaced by
$\mid\Psi(x_{0},Y)\mid^{2}$
in the path integral (13). In actual calculation,
the only difference is in the integration
over the field zero modes which now produces
$\delta(K-\sum_{j=1}^{M}p_{j})$.
The essential factor becomes then
$-\frac{1}{n(K^{2}-2)}$
for each $n$.
It follows that each mode $n$ contributes a probability measure
\begin{equation}
Z(\alpha)\prod_{q,i}e^{-\frac{1}{2}\alpha_{q}^{2}(1-\frac{K^{2}}{2})n}=Z(\alpha)\prod_{q,i}e^{-\frac{1}{2}\alpha_{q}(K)^{2}n^{2}},
\end{equation}
where $\alpha_{q}(K)^{2}=\alpha_{q}^{2}\frac{(1-\frac{K^{2}}{2})}{n}$.

Note that, since the $\alpha$ parameters
in both (17) and (18) are labelled by the momentum $K$, but {\it not} the
levels $m_{n}^{(i)}$, our results do not allow any
interpretation of the $\alpha$ in terms of
a quantum field on superspace, which is
dimensionally reducible to an
infinite tower of fields on space-time. For handles whose state is given
by (7), if the initial state is a state with definite values of the $\alpha$
parameters, the final state will be the same as the initial state [14],
according to our results, the radiation field $\alpha$ can
be dimensionally reduced to just one field, rather than a
tower of fields, on the usual space-time, i.e. on the prefered
minisuperspace from string-theory superspace, consisting of
just the $n=0$ modes [1,3]. Therefore, there will always exist
a set of classical values for $\alpha$ which makes it possible
to drive a consistent mechanism that fixes the values of
the coupling constants.

\vspace{0.5cm}

\noindent{\bf Acknowledgements.}
This work was supported by CAICYT under Research Project N§ PB91-0052.

\pagebreak

\noindent\section*{References}
\begin{description}
\item [1] S.W. Hawking, Nucl. Phys. B363, 117 (1991).
\item [2] A. Lyons and S.W. Hawking, Phys. Rev. D44, 3802 (1991).
\item [3] S.W. Hawking, Phys. Script. T36, 222 (1991).
\item [4] S.W. Hawking, Phys. Rev. D37, 904 (1988).
\item [5] S.W. Hawking, Nucl. Phys. B335, 155 (1990).
\item [6] S. Coleman, Nucl. Phys. B307, 867 (1988).
\item [7] A. Lyons, Ph. D. Thesis, University of Cambridge, UK, 1991.
\item [8] P.F. Gonz\'alez-D\'{\i}az, Nucl. Phys. B351, 767 (1991).
\item [9] S. Mandelstam, in {\it Unified Field Theories, Proceedings of
the 1985 Santa Barbara Workshop}, eds. M. Green and D. Gross (World
Scientific, Singapore, 1986).
\item [10] S.W. Hawking, in {\it 300 Years of Gravitation}, eds. S.W. Hawking
and W. Israel (Cambridge Univ. Press, Cambridge, 1987).
\item [11] W.H. Louisell, {\it Radiation and Noise in Quantum Electronics}
(McGraw Hill, New York, 1964).
\item [12] P.F. Gonz\'alez-D\'{\i}az, Mod. Phys. Lett. A8, 1089 (1993).
\item [13] P.F. Gonz\'alez-D\'{\i}az, Class. Quant. Grav. 10, 2505 (1993).
\item [14] S. Coleman, Nucl. Phys. B310, 643 (1988).

\end{description}

\end{document}